# Isotropic Dynamic Hierarchical Clustering

Victor Sadikov
at&t
200 S Laurel Av. A5-2D20
Middletown, NJ 07748
(732) 420-7453
vic@att.com

Oliver Rutishauser
Atlantic Style
P.O.Box 9
Oakhurst, NJ 07755
(732) 455-2081
rutishauser@yandex.com

*The only reasonable numbers are zero, one, and infinity.*

*Willem Louis van der Poel*

## ABSTRACT

We face a business need of discovering a pattern in locations of a great number of points in a high-dimensional space. We assume that there should be a certain structure, so that in some locations the points are close while in other locations the points are more dispersed. Our goal is to group the close points together. The process of grouping close objects is known under the name of **clustering**.

1. We are particularly interested in a **hierarchica**l structure. A plain structure may reduce the number of objects, but the data are still difficult to manage or present.

2. The classical technique suited for the task at hand is a **B-Tree**. The key properties of the B-Tree are that it is hierarchical and balanced, and it can be **dynamically** constructed from the input data. In these terms, B-Tree has certain advantages over other clustering algorithms, where the number of clusters needs to be defined *a priori*. The B-Tree approach allows to hope that the structure of input data will be well determine without any supervised learning.

3. The space is Euclidean and **isotropic**. This is the most challenging part of the project, because currently there are no B-Tree implementations processing indices in a symmetrical and isotropical way. Some known implementations are based on constructing compound asymmetrical indices from point coordinates, where the main index works as a key, while the function of other (999!) indices is lost; and the other known implementations split the nodes along the coordinate hyper-planes, sacrificing the isotropy of the original space. In the latter case the clusters become coordinate parallelepipeds, which is a rather artificial and unnecessary assumption. Our implementation of a B-Tree for a high-dimensional space is based directly on concepts of **factor analysis**.

4. We need to process a great deal of data, something like tens of millions of points in a thousand-dimensional space. The application has to be scalable, even though, technically, out task is not considered a true **Big Data** problem. We use dispersed data structures, and optimized algorithms. Ideally, a cluster should be an ellipsoid in a high-dimensional space, but such implementation would require to store $O(n^2)$ ellipse axes, which is impractical. So, we are using multi-dimensional balls defined by the centers and radii. On the other hand, calculation of statistical values like the mean and the average deviation, can be done in an **incremental** way. This mean that when adding a point to a tree, the statistical values for nodes of all levels may be recalculated in O(1) time. The node statistical values are used to split the overloaded nodes in an optimal way. We support both, brute force $O(2^n)$ and greedy $O(n^2)$ split algorithms. Statistical and aggregated node information also allows to manipulate (to search, to delete) aggregated sets of closely located points.

5. Hierarchical **information retrieval**. When searching, the user is provided with the highest appropriate nodes in the tree hierarchy, with the most important clusters emerging in the hierarchy automatically. Then, if interested, the user may navigate down the tree to more specific points.

The system is implemented as a library of Java classes representing Points in multi-dimensional space, Sets of points with aggregated statistical information (mean, standard deviation,) B-tree, and Nodes with a support of serialization and storage in a MySQL database.

## CCS Concepts

• Theory of computation➞ Theory and algorithms for application domains➞ Machine learning theory➞ Unsupervised learning and clustering • Mathematics of computing➞ Mathematical software➞ Statistical software • Information systems➞ Information retrieval➞ Retrieval tasks and goals➞ Clustering and classification

## 1. ( Points, Implementation )

In a high-dimensional space we assume that a considerable number of coordinates will contain zero values. To optimize the memory and storage space we would like to keep non-zero coordinates only. Thus, a Point object contains 3 fields: the number of non-zero coordinates, the array of sorted coordinate indices, and the array of corresponding coordinate values.

```
class Point
{
    int N = 0;       // # of non-zero coordinates
    int[] key;       // array of coordinate indices
    float[] val;     // array of coordinate values
}
```

The Point class provides methods for calculating the Euclidean length of the point vector, getting a particular coordinate (or zero,) adding another point to the given point, calculating a few useful functions like distance to another point, dot product, and serializing to a base-64 string.

Some functions, e.g. adding a point, may change the number of non-zero coordinates. The main loop for adding two points looks like the following.

```
int Ix = 0;  int Ip = 0;  int Iz = 0;
for(; Ix != this.N && Ip != p.N ;)
{
    if(this.key[Ix] == p.key[Ip])
    {
       z.key[Iz] = this.key[Ix];
       z.val[Iz] = this.val[Ix] + p.val[Ip];
       Ix++; Ip++; Iz++;
    }
    else if(this.key[Ix] < p.key[Ip])
    {
       z.key[Iz] = this.key[Ix];
       z.val[Iz] = this.val[Ix];
       Ix++; Iz++;
    }
    else
    {
       z.key[Iz] = p.key[Ip];
       z.val[Iz] = p.val[Ip];
       Ip++; Iz++;
    }
}
```

## 2. ( Sets, Mathematics )

The next step of our approach is the introduction of Sets of Points. The Sets allow calculation of aggregated statistical values. The most important value is the number of points (N) in the Set. It needs to be corrected each time a new point is added to the Set. The obvious way of calculating the new number of points is to increment the current number by one.

N = N + 1;

Other important statistical values of the set of points are arithmetic mean and the standard deviation. These values should also be adjusted every time a point is added to the Set. We could recalculate the arithmetical mean (**M**) from scratch, but we would like to follow the incremental approach and move it towards the newly added point (**P**) by the 1/N of the distance.

**M** = **M** + (**P** - **M**) /N;

As for standard deviation, at first sight, it seems to be a value that requires the full recalculation. Fortunately, this is not the case. We can store and adjust the standard deviation in an incremental way too, based on the following formula.

$E[X - E(X)]^2 = E[X^2] - (E[X])^2$

This means that to calculate the standard deviation it is enough to store the sum of the squares of point coordinates (S,) which can be adjusted incrementally.

S = S + |**P**|$^2$;

And when we need to calculate the standard deviation we will do the following.

D = sqrt( S/N - |**M**|$^2$ );

## 3. ( Clustering, Example )

Clustering basically means grouping similar objects together. If the objects have a number of numerical attributes they may be represented as points in a multi-dimensional space. The clustering will mean to partition the whole set of points into a number of disjoint sub-sets.

Let's consider an example in a one-dimensional space. The example is free of the challenges related to multi-dimensional clustering and is easy to comprehend. Assume we are given the set of five numbers { 0. 4. 5. 9. 13.}

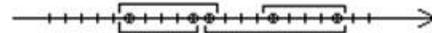

Figure 1

We can see that initially the points occupy the segment [0, 13.] Assume we need to split the set of the given points on the number line into two **clusters** then the resulting clusters will be sub-segments. Even in this simple example two different splits are possible. The first splits makes segments [0, 5] and [9, 13] while the other makes segments [0, 4] and [5, 13] (see Figure 1.)

In terms of segments, the first split looks better, because it finds two compact segments rich of points, while the second split covers almost the whole initial segment. In terms of statistical variables, the first split is better too, because the sum of the deviations of the resulting sets is minimal.

## 4. ( Dynamic Clustering, Approach )

If we continue adding new points, then we need to decide which sub-segment each new point belongs to. Points below 5 we can add to the first sub-segment, and points above 9 we add to the second sub-segment.

We may place points between 5 and 9 into either sub-segment. Our criterion here is to avoid big segments, or (which is the same) to keep the sum of deviations minimal. But we always need to update the boundaries of the sub-segments so that we can exactly know in which segment a particular point is to be found. If the points are well spread, knowing exact boundaries of the segments may also improve unsuccessful search.

After adding a certain number of points to a sub-segment, we will need to split this sub-segment into two sub-segments with a smaller number of points. This can be done exactly in the same manner as we split the initial segment. After that we will be adding new points to the set of three sub-segments. Then we will need to split another sub-segment. And the number of sub-segments will increase again.

Unlike the static approach where the set of all objects exists before the procedure of clustering starts; dynamic approach assumes that clusters are incrementally adjusted each time a new object gets added into the set. This eliminates the dedicated step of clustering for the price of a longer time needed to include objects. Dynamic clustering provides better flexibility and can be performed with less *a priori* known information about the data, e.g. when the total number of target clusters is unknown.

## 5. ( Hierarchical Clustering, Approach )

The bigger the number of points in our data set, the bigger the number segments. Soon it gets big enough, and we may need to introduce the next level of **hierarchy**, when smaller clusters are, in their turn, grouped into clusters of the higher level. The approach of adding points, splitting segments, and adding new levels when necessary is quite similar to adding objects to a B-tree [1.] The tree starts with the root node, responsible for all points stored in the tree. At each level, the parent node consist of a number of sub-nodes. The points in each sub-node are close one to another.

A classical one-dimensional B-tree design focuses on minimization of the information stored at the node level. Namely, a parent node stores a number of adjacent values in ascending order, with the sub-nodes being placed between adjacent values, plus one at the beginning and one at each end. Thus we know that the elements of each sub-tree are greater than the left adjacent value and less than the right one.

On the contrary, our design prefers to store excessive boundary and statistical information about the sub-nodes. In the case of one-dimensional space, each node occupies a segment and sub-nodes of one parent do not intersect. The boundary segment can be defined by its two ends, or by the middle point (**C**) and the distance (R) to the ends. The letter way will occupy less memory in a general multi-dimensional case. We also keep the number of points in each sub-tree, their arithmetic mean and standard deviation. As we stated above, statistical values facilitate splitting the nodes in the optimal way, while the boundary information allows us in some cases to terminate an unsuccessful search heuristically.

## 6. ( Isotropic B-tree, Approach )

B-trees do their work great, as long as the attributes of the objects are one-dimensional. Unfortunately, there are some problem with direct extension of B-tree to a multi-dimensional case. The two common approaches are the following.

The mix approach assumes composing the compound index based on the component indices, and then making use of a one-dimensional B-tree. The problem with this approach is that the component indices are treated by far not equally. One index plays the main role, while the role of the others is insignificant.

The other approach is more complicated. R-tree [2] is a variant of B-tree where the nodes are bound by coordinate rectangles. This approach is more symmetrical in terms of using indices. But the directions along the coordinate axes are still different from arbitrary ones.

We would like to build a variant of B-tree, where the nodes are bound with circles or ellipses. This decision ensures that the essential property of isotropy of the physical space is not ignored.

## 7. ( Isotropic Dynamic Hierarchical Clustering, Two-Dimensional Case )

The circles are defined by the center point (**C**) and the radius (R.) It the 2-dimensional space, the center is defined by the two coordinates, and so we need to store 3 real values. Alternatively, if we decided to present clusters as ellipses in general orientation, we will need to store the semi-principal axes and the angles, which would require $O(n^2)$ memory, with *n* being the number of dimensions in the space.

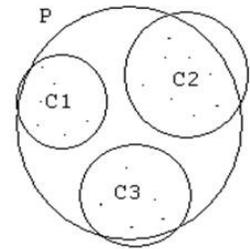

In the picture to the right, the circles corresponding to the sub-nodes of one and the same parent do not intersect. The biggest circle (P) corresponds to the parent node, while $C_1$, $C_2$, and $C_3$ correspond to the sub-nodes. Keeping on adding new points to the tree we cannot avoid the situation where the sub-nodes start to intersect. We will dwell on the intersecting areas later in this paper.

### 7.1 Selecting a Sub-Node

As mentioned in section 4, while adding a new point outside any bounding circle, we need to select the most appropriate sub-node. To do this, we will try to add the new point to each sub-node, and calculate the new bounding radii. Then we will select a sub-node so that the sum of the new squared radii should be minimal, because our goal is to make eventually all sub-node circles of about the same absolute size in the given space.

Let's assume that the old radius was $R_i$, then the new radius will be $R_i+H_i/2$, where $H_i$ is the distance from the new point to the circle $C_i$. If we select the i-th sub-node, the sum of new squared radii will grow by $(R_i+H_i/2)^2 - R_i^2$, i.e. by $R_iH_i+H_i^2/4$, so we need to select the sub-node where $T_i = 4R_iH_i+H_i^2$ is minimal.

### 7.2 Splitting the Overloaded Node

When a node gets overloaded, i.e. it includes more than the maximum number of sub-nodes or points; the node needs to be split into two nodes at the same level. In the classical

B-tree the node is split into two nodes with the equal number of elements.

In our case, we can split the node taking into account the following criteria:

• the maximum radius of the two new circles is minimal;
• the new nodes intersect with the minimal area;
• the sum of standard deviations of the new nodes is minimal.

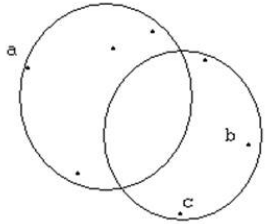

To this goal, first we will find the pair of points at the *longest* distance. In the figure to the right, such a pair consists of points **a** and **b**. If these points both belonged to one and the same bounding circle, the radius of this circle would be greater than the distance $D_{a,b}/2$, which, as we assume, is the maximum distance. So, to minimize the maximum radius we need to distribute points **a** and **b** to the different bounding circles $C_a$ and $C_b$. Now we will find point **c**, so that the distance from **c** to either of circles $C_a$ and $C_b$ is the longest. In the picture above, it is the distance between point **c** and circle $C_a$ (which now consists of just one point **a**.) Once again, to minimize the would-be radii, we need to distribute point **c** to the other circle, $C_b$. Continuing in the same manner, we will ultimately get circles $C_a$ and $C_b$, as shown in the picture.

### 7.3 Adjusting Bounding Circles

Let's assume that we need to add a new point (**Q**) to the tree. We start from the root and go down the tree, level by level. At each level we need to select the most appropriate sub-node. E.g. at the parent level P, we need to select one of the sub-nodes, $C_1$, $C_2$, or $C_3$. Logically, there are two different cases. If the new point belongs to a particular bounding circle, no adjustment is needed. But if the new point is outside of any bounding circle, we need to select the most appropriate sub-node; to add the new point to the selected sub-node; and to adjust the corresponding bounding circle, so that it should include the new point as well as all old points. See the figure below.

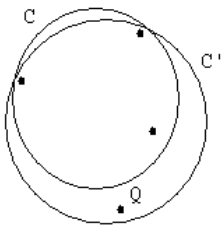

In the 2-dimensional case, the new minimal bounding circle can be exactly calculated. In the figure above, the minimal circle is the circle circumscribed about the triangle of the first two points and point **Q**. It can be calculated from the coordinates of these points. E.g. the radius of the circumscribed circle is $L_1*L_2*L_3/\sqrt{(L_1+L_2+L_3)*(L_2+L_3-L_1)*(L_1-L_2+L_3)*(L_1+L_2-L_3)}$ ), where $L_1$, $L_2$ and $L_3$ are the lengths of the sides.

In a multi-dimensional space, the situation is much more complicated. Fortunately, we do not need the exact minimal bounding circles. The bounding circles are very useful in many procedures where they heuristically allow to reduce the amount of calculation, but fortunately they are not critical. So, we would recommend to use a less exact but easier to calculate approximation.

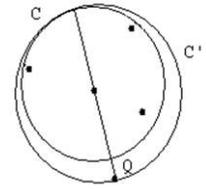

We can easily construct the new bounding circle (C') about the old bounding circle (C) and the newly added point (**Q**) as shown in the figure to the right. First, we calculate the distance (H) from the point **Q** to the circle C. Then we move the center of the circle C towards **Q** by H/2. This will be the center of C'. The radius of the new circle will be the old radius (R) plus H/2. Thus, the circle C' will surely contain all old points as well as the new point.

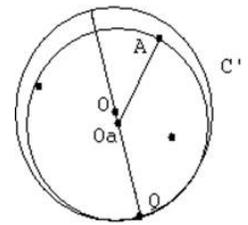

Moreover, the new bounding circle (C') can be easily optimized. By construction, the new circle has to lie through the point **Q**, but it doesn't probably lie through any other actual points. So, we can shrink the circle, so that it lies through yet another point.

Let's shift the center (**O**) of the circle towards the point **Q**, so that the new circle with the center **Oa** lies both through points **Q** and **A**. See the picture above. The value of the shift can be calculated based on the points **Q**, **O** and **A**. When we shift the center for the point **A**, this doesn't necessarily mean that any other point **B** will belong to the new shrunken circle. But we can repeat this procedure for all points of the set, and find the largest shrunken circle, corresponding to the shortest allowed shift. That circle will work for all the points of the set.

### 7.4 Calculation of a Quasi-Minimal Bounding Circle

Let's assume that point **T** is the foot of the perpendicular dropped from point **A** on the line (**Q**,**O**.)

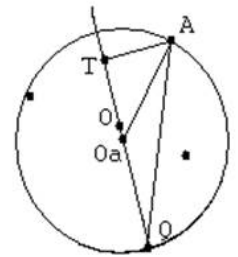

Then $|A\ T|^2 + |T\ Q|^2 = |A\ Q|^2$. and $|A\ T|^2 + |T\ Oa|^2 = |A\ Oa|^2$.

But $|T\ Oa| = |T\ Q| - |Oa\ Q|$ and $|Oa\ Q| = |Oa\ A|$. So, finally,

$|Oa\ Q|$ should be $|A\ Q|^2 / 2\ |T\ Q|$, where $|T\ Q|$ is the projection of the vector $[Q\ A]$ on line (**Q**,**O**) and can be calculated by means of the dot product.

### 7.5 Exact Minimal Bounding Circle

Please notice that, in some cases, the exact minimal bounding circle may be slightly smaller than the quasi-minimal circle constructed above. Moreover, the shrunken circle (C') depends on the initial circle C.

If the set consists of just one point, the exact minimal circle is the circle with the center in that point and the radius of zero. If the set contains two points, the exact minimal circle is the circle from the center of mass and the radius of half the distance between the points.

If the set contains three points, there are two cases. Either, the exact minimal circle is the circle circumscribed around these three points. Or, the circle built on the two of the three points as a diameter, provided that the third point lies inside it. If we want to build the circle for the second case, we need first to find the two (out of the three) points with the longest distance between them. Then, we need to check that the third point (**X**) will make an obtuse-angled triangle. I.e. $|A\ X|^2 + |X\ B|^2 < |A\ B|^2$. Sets of more than three points are quite similar.

If we do not want to build the exact minimal circle, we may use the shrinking technique. But we will need a good first approximation for the minimal circle. As shown above, the circle build on the longest segment plays an essential role in construction of minimal circles; besides, and it is easy to calculate. This makes it a good first approximation for building quasi-minimal bounding areas.

## 8. ( Multi-Dimensional S-Tree, Full Details)

So far we have clearly described what we would like to achieve in our multi-dimensional S-Tree. A real implementation is not so smooth, and requires solutions to a number of complicated issues.

### 8.1 Overlapped Circle

As mentioned in Section 7, some sub-nodes of a given node may overlap. Such cases may occur when a new point is added to a sub-node, which results in adjusting the bounding circle of this sub-node, or when a sub-node is split into two sub-nodes at the same level, as discussed in Section 7.2. In this case the actual areas of sub-nodes are not circle, but they are rather "cut" circles, as shown in the picture below.

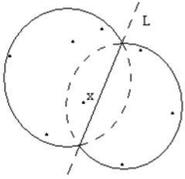

The reason for the sub-node areas to be nothing but the "cut" circles is that the areas need to be convex. Now, the more complicated form of the areas makes us change the way we calculate the appropriate sub-node when the given point **x** belongs to both bounding circles.

In the two-dimensional space we use line L, to determine where point **x** belongs to. All points at one side of line L belong to one sub-node, while all the points at the other side belong to the other sub-node. Analogically, in the multi-dimensional space, the border line L will become a plane. All points at one side of the plane will belong to one sub-node, and all points at the other side of the plane L will belong to the other sub-node.

Every plane in the multi-dimensional space can be defined by its normal vector and the distance from the point of origin. The only drawback is that theoretically we will need a border plane for each pair of sub-nodes. E.g. if a parent node has, say, 5 sub-nodes, we will need to store 10 border plains. A more memory-effective way would be to calculate the equation of the plane L on the fly, based on the bounding circles.

Namely, as we discussed, the circle $C_1$ can be defined by its center, $O_1$, and the radius $R_1$. Analogically, we define the circle $C_2$ by the center $O_2$ and the radius $R_2$. It is easy to see that the border plane L is the set of all the point such as the difference between their squared distances to points $O_1$ and $O_2$ is constant.

$|O_1 - x|^2 - |O_2 - x|^2 = F_{1,2}$

To be precise, the constant $F_{1,2}$ is the difference between the squared radii of the circles in question, i.e. $F_{1,2} = R_1^2 - R_2^2$

So, to find whether point **x** belongs to the circle $C_1$, we need to calculate $|O_1 - x|^2 - |O_2 - x|^2 - R_1^2 + R_2^2$, and to compare it with zero.

### 8.2 Splitting the Node

When a node gets overloaded, i.e. it includes more than the maximum number of sub-nodes or points; the node needs to be split into two nodes at the same level.

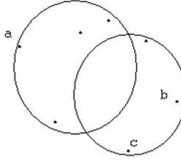

While splitting the node, we expect that the results nodes will have bounding areas with smaller radii. Let's assumes that **a** and **b** are the points with the greatest distance between them, as in the picture above. If we ultimately put these points into one and the same node, the radius of the bounding circle for that node can not be less than |**a b**|, which is almost the radius of the bounding circle for the original node. To avoid this, we have to put **a** and **b** into different nodes. Now we have nodes Ca and Cb, consists of points **a** and **b**, correspondingly.

At the next step let's consider, say, the point **c**. We can either put it into Ca or Cb. And we need to estimate how good or bad it would be to put it into a particular node. E.g. we can try to optimize (to keep minimal) the maximum radius of Ca and Cb. But, now we know that there is something more unpleasant than just big radii; it is overlapped circles. So we may want to keep the circles overlapped in the minimal possible measure.

In the previous section we have introduced the expression $L^2 – (R^2 – r^2)$. It shows in what manner the circles are overlapped, and should be greater than zero. We will try to optimize (to keep maximal) this expression. It gives the same rules for selecting points **a**, **b**, and **c**, but gives better results at next steps.

## 9. ( Example, Two-Dimensional Case )

The most challenging task of implementing a B-tree is a split of overloaded nodes. When a node is split into two nodes, there appears a new boundary. All sub-nodes of the given node may need to be recursively split by the new boundary. The picture below illustrates the result of a split for a two-dimensional case.

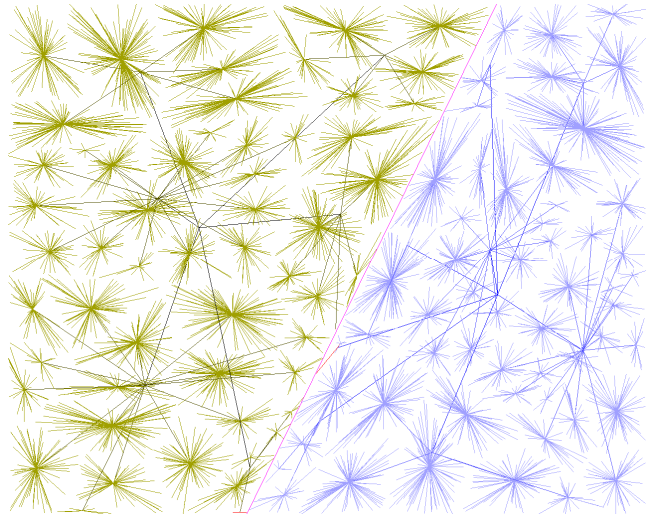

Here the new boundary is highlighted in red. The new sub-trees are painted olive and violet. The higher levels are darker than lower levels, so that it would be easier to trace the centers of clusters.

## 10. ( Illustration, Three-Dimensional Case )

As we discussed in Section 8.1, ideally, the clusters of a particular level should form isotropic ellipsoids. The result we would like to get should look like soap foam bubbles.

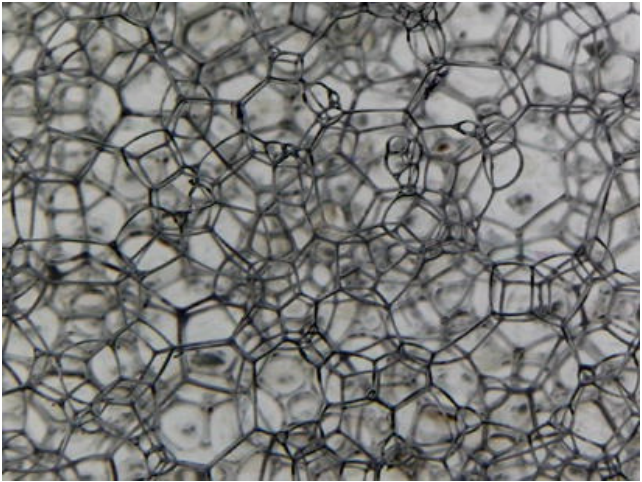

## 11. ( Example, Multi-Dimensional Case )

As an example we will cluster wiki pages.

1. The first step is to parse an HTML page and to extract pure text. There are several tools available for this purpose. We use JavaCC, an open source parser and lexical analyzer generator. The generator accepts a formal grammar definition, written in .JJ file, which also allows to define additional custom code. The input to the lexical analyzer is a sequence of characters; the output is a sequence of tokens.

In our example, we are interested in skipping HTML tags and parsing the text further, so that it become a list of words. At this stage we are dropping everything what is not a word, i.e. numbers, email addresses, references to web pages, expressions, identifiers, etc.

2. Then, we analyze the text and map it to a point in a semantic space. For each word in text we will find the root. Basically, for verbs we drop endings as -s, -ed, -ing; for nouns we drop ending –s. Actually, the procedure is a bit more complicated due to language exceptions. Secondly, we calculate the weight of the word. We assume that more frequent words should have a lighter weight than infrequent words. So, we distributed all the words to 256 sets with about equal frequencies. Thirdly, we find the meaning of the word in question. We have split all words to 1024 groups with similar meanings. Please notice, that one word can have more than one meanings, with only one of them being actualized in the text. Without knowing what the actual meaning is, we have to add all meaning with the same weights depending on the frequency of the word.

Now we can define a target point in a 1024-dimensional space where each dimension corresponds to a meaning. The coordinates of the target point are calculated by accumulating all weights corresponding to particular meanings. It also seems reasonable to divide the coordinates by the number of the words in the text, so that repletion of sentences or words does not affect the meaning of the text.

## 12. CONCLUSIONS

Arbitrary points in multi-dimensional space can be isotropically clustered into a balanced hierarchical structure, similar to a B-tree.

Clustering into a multi-dimensional B-tree does not require any supervision or any *a priori* given information, like the number of clusters.

Clustering into a multi-dimensional tree can be done dynamically and efficiently. Adding new points to the tree requires only incremental updates of statistical values associated with nodes.

Text pages can be mapped to points into 1000-dimentional semantic space.

The search of pages close to a given semantic point can return a hierarchically ordered results, allowing the user to select more general or more specific topics.

## 13. ACKNOWLEDGMENTS

This research was not sponsored by National Science Foundation or any other financial source.

## 14. REFERENCES

[1] Donald E. Knuth, 1973, *Sorting and Searching*, volume 3, *The Art of Computer Programming*, Addison-Wesley.

[2] Antonin Guttman, 1984, *R-trees: a dynamic index structure for spatial searching,* ACM, New York, USA. (1984-r-tree-guttman.pdf)

[3] Sean Owen et al. 2011, *Mahout in Action,* Manning Publications Co., New York, USA. (ISBN 9781935182689, Mahout.in.Action.pdf)

[4] Richard A. Reyment, K. G. Joereskog, 1993, *Applied Factor Analysis in the Natural Sciences*, Cambridge University Press, UK.

[5] George A. Miller, 2003, *WordNet Lexical Database of English Language*, Cognitive Science Laboratory of Princeton University

[6] Roget's Thesaurus, 2006, Electronic Lexical Knowledge Base (ELKB) http://www.nzdl.org/ELKB

[7] Adam Kilgarriff, 1995, *BNC Database and Word Frequency Lists*, http://www.kilgarriff.co.uk/bnc-readme.html

[8] Brian Goetz's, 2003, *HTML Parser.*